\documentclass[
]{ceurart}

\usepackage{listings}
\usepackage{enumitem}
\usepackage{xurl}
\ExplSyntaxOn
\NewDocumentEnvironment{resultfigure}{}{
  \par\addvspace{10pt}\noindent
  \begin{minipage}{\linewidth}
  \__reset_fig:
  \bool_gset_false:N \g_fig_full_bool
  \dim_set:Nn \l_fig_width_dim {\linewidth}
  \cs_set:cpn {@captype} {figure}
  \cs_set_eq:cN {@makecaption} \__make_fig_caption:nn
  \centering\sffamily\small
}{
  \end{minipage}\par\addvspace{10pt}
}
\ExplSyntaxOff

\begin{document}

\copyrightyear{2026}
\copyrightclause{Copyright for this paper by its authors.
  Use permitted under Creative Commons License Attribution 4.0
  International (CC BY 4.0).}

\conference{ITASEC \& SERICS 2027: Joint National Conference on Cybersecurity,
  February 08--12, 2027, Siena, Italy}

\title{Rapid Mobilisation and Organisational Fragmentation: The Telegram Ecosystem during the 2026 Italian Student Protests}

\author[1,2]{Lorenzo Alvisi}[%
orcid=0009-0007-4222-348X,
email=lorenzo.alvisi@imtlucca.it,
]
\cormark[1]
\address[1]{IMT School for Advanced Studies Lucca, Piazza San Francesco 19,
  55100 Lucca, Italy}
\address[2]{Institute of Informatics and Telematics, National Research
  Council (IIT-CNR), Via G. Moruzzi 1, 56124 Pisa, Italy}

\author[2]{Maurizio Tesconi}[%
orcid=0000-0001-8228-7807,
email=maurizio.tesconi@iit.cnr.it,
]

\cortext[1]{Corresponding author.}

\begin{abstract}
In October 2026, social media pages named \emph{Italia in Piazza} used several platforms to call for demonstrations in Italian cities. The mobilisation brought participants into a rapidly assembled network of Telegram chats, but disagreements over its direction and protest tactics became visible after the first demonstrations. We study how this online environment formed and subsequently fragmented. We first constructed an ecosystem of chats connected to the protests, then compared it with established conspiracy, radical-left and radical-right ecosystems. Few links or observed participants connected the protest ecosystem to these communities. Its messages nevertheless expressed a broader mix of political positions, brought together by a shared focus on the protests. Within this ecosystem, geographically distributed discussions were linked through a small administrative core and connections to TikTok and Instagram. The mobilisation subsequently developed into two distinct spaces: Movimento 17 Ottobre and a re-formed Italia in Piazza chat. The former contained more content classified as promoting violence or hate, while the latter retained part of the original staff and advocated organised, non-violent protest. Their observed populations showed little overlap. After the original chats became inaccessible, discussion was rebuilt on Telegram, while Movimento 17 Ottobre encouraged migration to Signal. The case connects the fragmentation of a shared mobilisation with questions of deplatforming and community governance: the loss of discussion spaces need not end collective activity, while the environments that emerge can develop markedly different norms around protest and violence.
\end{abstract}


\maketitle

\section{Introduction}

In autumn 2026, youth-led protests spread across Europe. In France, Belgium, Spain, Portugal, Albania and Italy, students and young workers mobilised against precarious employment, rising living costs, housing shortages and underfunded public schools \cite{albero2026giovani}. In Italy, high-school students explicitly drew inspiration from the French mobilisation. On 16 October, student marches in several cities combined calls for school funding with opposition to rearmament and rising living costs \cite{convertino2026studenti}.

On 17 October, demonstrations against rising living costs and low wages took place in several Italian cities. The call to protest circulated across social media platforms under the name \emph{Italia in Piazza} \cite{ilpost2026scontri,ilpost2026gruppi}. Its Instagram page opened on 4 October, while mobilisation developed through TikTok and a WhatsApp community. When the latter reached its membership limit, participants moved to Telegram \cite{dimalio2026alleanza}. We refer to the original national Telegram chat as \textbf{IIP}. Around it, we identified a network of 19 chats linked through six core administrators: IIP, 16 regional groups and two local groups. We call this the \emph{Italia in Piazza chat network}, distinguishing it from the broader set of chats discussing the protests (Section~\ref{sec:admins}).

Thousands took part in the demonstrations. In Milan, Turin and Bologna, clashes with police and looting followed the marches. According to press reports, those involved in the clashes were not affiliated with established collectives or social centres \cite{ilpost2026scontri,skytg24_2026scontri,ilpost2026bologna}. Unrest continued in central Milan on 18 October, when hundreds of protesters were involved in further clashes and looting, and more than 200 people were taken to police headquarters \cite{ilpost2026domenica,skytg24_2026milano}.

\begin{figure}[t]
  \centering
  \includegraphics[width=\textwidth]{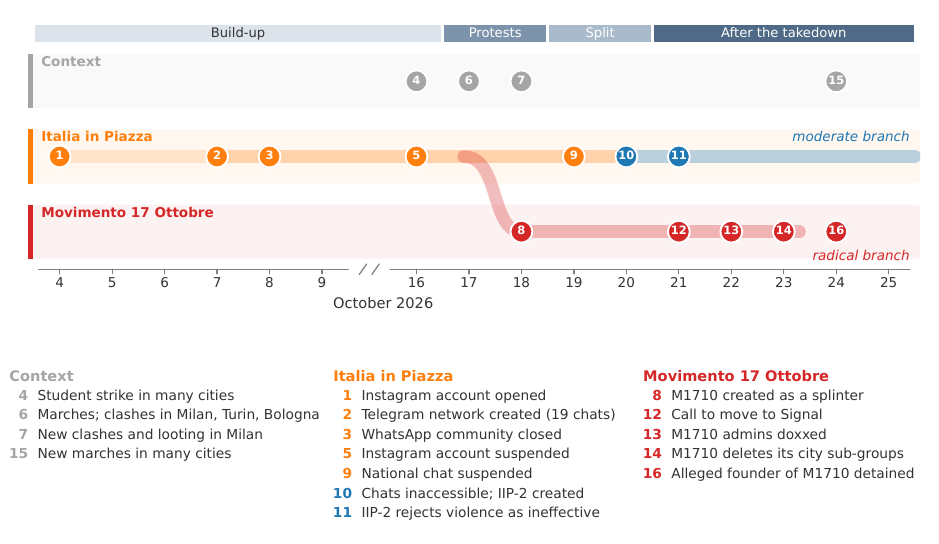}
  \caption{Timeline of the mobilisation, 4--25 October 2026. Each lane follows one actor, the ribbon at the top marks the phases of the mobilisation and the numbered markers refer to the events listed below the plot by lane. The days from 10 to 15 October, without events, are compressed (//). The thick bars show when each chat was active: the WhatsApp community and the Italia in Piazza chat network form a single trunk, which splits after the protests into a more radical branch, M1710, and a more moderate branch, IIP-2. The timeline extends to 25 October, the end of IIP-2's collection and the date by which the original Italia in Piazza founder's account had been deleted; the protest-ecosystem dataset (Table~\ref{tab:datasets}) ends on 23 October, the date of M1710's last message.}
  \label{fig:timeline}
\end{figure}

The aftermath exposed two contrasting currents within the mobilisation: a more radical component and a more moderate one. Two new Telegram spaces gave these currents distinct organisational settings. On 18 October, \emph{Movimento 17 Ottobre} emerged as a splinter movement, accusing Italia in Piazza of tolerating racism and fascism \cite{dimalio2026alleanza} and announcing a ``permanent mobilisation'' every Saturday \cite{ilpost2026gruppi,dimalio2026telegram}. We call its Telegram chat \textbf{M1710}. On 20 October, IIP and the original regional chats became inaccessible. Administrators attributed the closure to a Telegram takedown and established a new national chat within hours, with substantial involvement from the original regional administrators. We call this re-formed chat \textbf{IIP-2}. M1710 became the more radical space, while IIP-2 articulated a more moderate approach to protest. On 24 October, amid further demonstrations \cite{domani2026proteste}, the Milan prosecutor's office ordered the detention of a man whom investigators identified as the founder of Movimento 17 Ottobre. Investigators alleged incitement to commit crimes aggravated by terrorist purposes and identified him and his wife as the sole administrators of its Telegram chat and Instagram account \cite{skytg24_2026fermo,ansa2026amministratori}. We report this account as context for the mobilisation's chronology (shown in Figure~\ref{fig:timeline}), without independently verifying these administrative roles or assessing criminal responsibility.

To compare this mobilisation with other thematic communities on Italian Telegram, we use \emph{ecosystem} to mean a cluster of chats sharing a topic or thematic focus. An ecosystem can contain distinct groups and channels, with different participant populations, political positions and communication practices; a shared topic does not imply a homogeneous audience or common leadership. The Italia in Piazza chat network is therefore a subset of the broader \emph{protest ecosystem}, which also includes other chats connected to the mobilisation. To construct the comparison ecosystems, we selected seed chats based on our prior knowledge of Italian Telegram communities gained through the study reported in~\cite{alvisi2026mapping}. We then expanded each seed set through the forward-based snowball procedure and manual relevance checks described in Section~\ref{sec:collection}. We consider the \emph{conspiracy} ecosystem, which developed around opposition to COVID-19 vaccination and later shifted towards pro-Russian geopolitical narratives; the \emph{radical-left} ecosystem, including social centres, student and climate collectives and pro-Palestine groups; and the \emph{radical-right} ecosystem, including neo-fascist parties and their student organisations. These provide contrasting settings in which opposition to government and street mobilisation can be expressed.

The study proceeds from a comparative characterisation of these ecosystems to an examination of the protest movement's internal split. First, we ask how the ecosystems are structured and which populations they bring together, distinguishing network organisation, audience composition and expressed political positions. We then examine how the two currents that became visible after the protests developed into separate spaces, comparing the more radical M1710 with the more moderate IIP-2. We address two research questions:
\begin{enumerate}[label=\textbf{RQ\arabic*.}, leftmargin=*]
  \item How do the protest, conspiracy, radical-left and radical-right Telegram ecosystems differ in their network structure, cross-platform connections, participant populations and expressed political positions, and where does the Italia in Piazza chat network sit within this landscape?
  \item After the radical and moderate currents of the mobilisation became visible, how did M1710 and IIP-2 differ in leadership continuity, audience composition and attitudes towards violence as a form of protest?
\end{enumerate}

To answer RQ1, we reconstruct the ecosystems through a Telegram crawler that collects messages, resolves forwards and outbound links, and represents connections among chats and to accounts on other platforms as a multilayer network. We combine structural measures with audience overlap and LLM-based annotation of expressed political orientation and stance towards the protests. These complementary perspectives describe both the infrastructure of each ecosystem and the differences among its participating populations, without treating an ecosystem as a single political actor. To answer RQ2, we trace administrators and observed participants from the original chat network into M1710 and IIP-2, comparing leadership continuity, audience composition, cross-platform connections and content promoting violence or hate. Preliminary findings were reported in the Italian press \cite{dimalio2026admin}.

After the split, the two chats took different directions. IIP-2 retained many of the original administrators and encouraged participation in organised protests, including a union-called general strike. M1710, by contrast, hosted more content promoting violence and hate and ceased posting within five days. Using the same annotation procedure for both chats, we identified such content in 1.7\% of M1710 messages and 0.14\% of IIP-2 messages. These differences concern the content shared in each chat and do not imply that all participants endorsed it.

The split also matters for platform governance: chats emerging from the same mobilisation differed substantially in the prevalence of content promoting violence and hate. Examining these differences alongside leadership continuity and audience composition helps place potentially harmful content in its organisational context, rather than treating the protest network as a single, uniform community.

\section{Related Work}
\label{sec:related}

Telegram brings together public channels that distribute content and groups in which participants discuss and organise. Forwarded messages and shared links connect these spaces, allowing researchers to reconstruct networks of related communities. Large-scale collections such as TGDataset cover extensive sets of public channels~\cite{lamorgia2023tgdataset}, while studies of far-right and conspiracy communities have used snowball sampling to follow connections from selected seed chats~\cite{bovet2022organization,urman2022they,hoseini2023globalization}. Forward-based collection has also supported comparisons of Italian and English-speaking conspiracy communities~\cite{alvisi2024unraveling} and the mapping of thematic communities across the broader Italian Telegram landscape~\cite{alvisi2026mapping}.

These networks reveal connections that a classification by topic alone would miss. Communities can discuss similar issues while drawing on different sources, expressing different political positions and circulating different narratives. During the COVID-19 pandemic in Italy, for example, online discussion ranged from fear about the outbreak to opposition to public-health measures, institutional distrust and street mobilisation~\cite{calamusa2020twitter,spitale2022concerns}. Research on conspiracy-oriented Telegram discussions also documented hate speech~\cite{vergani2022hate}, while the broader mapping of Italian Telegram examined thematic communities alongside political orientation and harmful content~\cite{alvisi2026mapping}.

Differences also exist within individual communities. Users may adapt their language to local norms, and the tone of messages does not necessarily match the audience reactions they receive~\cite{alvisi2026toxicity,tardelli2026emoji,avalle2024persistent}. A chat's most visible content therefore offers only a partial account of its participants. LLM-assisted text analysis provides additional tools for studying the content of online ecosystems~\cite{alvisi2026mapping,popa2025personality,petrillo2025tallm,senette2024inferring}, alongside analysis of their connections and audiences.

When these communities participate in a common protest, their differences need not be resolved before mobilisation begins. Digital platforms allow personalised calls to action to circulate through loosely connected networks, enabling participation without membership in a formal organisation~\cite{bennett2012logic,earl2011digitally}. Telegram's role in the Hong Kong and Belarus protests illustrates how channels and groups can support the circulation of protest information and coordination~\cite{urman2021analyzing,wijermars2022telegram}. In Italy, Telegram also provided organising spaces for protests against COVID-19 restrictions~\cite{ap2021bastaDittatura}.

The ability to mobilise, however, does not necessarily equip participants to make collective decisions or manage disagreement. Tufekci~\cite{tufekci2017twitter} describes how digitally enabled movements can grow rapidly without developing the organisational capacity to negotiate goals and adapt tactics. The \emph{gilets jaunes} similarly combined online activism with street action while mobilising without institutional or third-party support~\cite{royall2020gilets}. Participation in the same mobilisation can thus precede agreement on who represents it or how it should act.

The content circulating through these networks can also have security implications: disinformation has been studied as a means of targeting people and critical infrastructures~\cite{alvisi2024weaponizing}. The spaces sustaining mobilisation are subject to platform rules, account restrictions and content removals~\cite{gillespie2018custodians}. Closing a discussion space may disperse its participants, but it may also prompt them to regroup elsewhere: research on deplatforming has documented migration to Telegram and other services~\cite{rogers2020deplatforming}. The outcomes vary. Migration can reduce activity while concentrating harmful content among those who remain~\cite{hortaRibeiro2021platformMigrations}, whereas research on Reddit bans found reductions in hateful behaviour among users who continued using the platform~\cite{chandrasekharan2017you}. These contrasting outcomes complicate the assessment of moderation interventions, alongside the difficulties of interpreting content in context when moderation operates automatically and at scale~\cite{gorwa2020algorithmic}.

\section{Methodology}
\label{sec:method}

\subsection{Data collection}
\label{sec:collection}

We collected Telegram data using a crawler built with the Telethon library. Throughout the study, we use \emph{chat} to refer to both groups, where members can contribute, and channels, where posting is restricted to administrators.

Collection started from a set of seed chats for each ecosystem. For the protest ecosystem, these were chats associated with Italia in Piazza. For the comparison ecosystems, seed chats were selected from communities identified in the study reported in~\cite{alvisi2026mapping}. We then expanded each set using a forward-based snowball procedure~\cite{alvisi2024unraveling,alvisi2026mapping}.

At each iteration, the crawler ranked referenced chats using three criteria in sequence: whether the researcher account had already joined them, the number of ecosystem chats referencing them, and the total number of references. We manually inspected candidates and added those relevant to the ecosystem's topic. Expansion stopped when the highest-ranked candidates no longer matched that topic. We retained forwards and links between ecosystems, allowing their connections to remain visible in the collected data.

For each chat, the crawler retrieved messages, forwarded content and its source, and outbound links. It resolved Telegram usernames, invitation links and private-chat links, as well as links to accounts on TikTok, Instagram, X, YouTube and Facebook. For shared TikTok videos, it also retrieved descriptions and up to 100 comments per video, without comment-author identifiers. Before analysis, we removed service messages, messages posted by or through bots, and verbatim duplicates.

\subsection{Dataset overview}
\label{sec:datasets}
\label{sec:admins}

The dataset comprises four Telegram ecosystems: protest, conspiracy, radical left and radical right. The protest ecosystem includes chats associated with Italia in Piazza and others reached through the expansion procedure, including M1710, local protest chats and chats recruiting participants within the mobilisation. The comparison ecosystems cover conspiracy communities, radical-left collectives and radical-right organisations. Three chats reached during expansion of the protest ecosystem were assigned to the radical-left dataset.

Within the protest ecosystem, we identify the \emph{Italia in Piazza chat network} through its association with a shared administrative core. When users asked who led Italia in Piazza, the Group Help moderation bot in IIP returned six accounts: a founder, a vice-founder and four staff members. We refer to these six accounts collectively as the \emph{staff}, following their presentation in the chat.

\begin{figure}[t]
  \centering
  \includegraphics[width=0.8\textwidth]{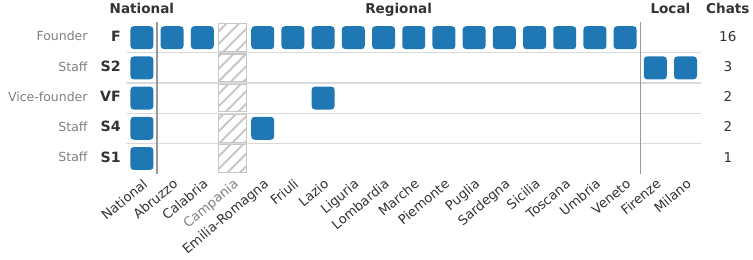}
  \caption{Italia in Piazza chats administered by each staff account, from lists saved on 19 and 23 October. S3 is absent from these lists; Campania's administrator list was inaccessible (hatched column). F: founder; VF: vice-founder; S1--S4: staff members.}
  \label{fig:admins}
\end{figure}

The \emph{Italia in Piazza chat network} comprises 19 chats associated with these staff accounts: IIP, 16 regional chats and two local chats for Firenze and Milano. The regional chats were named after their respective regions, shared IIP's profile picture and were explicitly presented as regional groups on the main Italia in Piazza page. These features supplemented the administrator lists in establishing their association with the network (Figure~\ref{fig:admins}). Other local protest chats, including Udine and Naples, had separate creators and administrators with no staff accounts in common. They belong to the protest ecosystem but are not included in the \emph{Italia in Piazza chat network}.

Table~\ref{tab:datasets} summarises the four datasets. The protest ecosystem contains 38 chats, compared with 25 in the conspiracy ecosystem, 20 in the radical-left ecosystem and 40 in the radical-right ecosystem. Groups account for 29 of the 38 protest chats, including all 19 chats in the \emph{Italia in Piazza chat network}, whereas channels predominate in the comparison ecosystems.

\begin{table}[t]
  \centering
  \small
  \setlength{\tabcolsep}{4pt}
  \begin{tabular}{lrlrrrr}
    \toprule
    Ecosystem & Chats & Period (2026) & Messages & Forwards & \texttt{t.me} links & Social links \\
    \midrule
    Protest                & 38 & 20 Sep -- 23 Oct & 85{,}055 & 733   & 1{,}214 & 446 \\
    Conspiracy             & 25 & 21 Sep -- 21 Oct & 65{,}533 & 6{,}769 & 8{,}306 & 3{,}777 \\
    Radical left           & 20 & 19 Sep -- 19 Oct & 681      & 188   & 43    & 14    \\
    Radical right          & 40 & 20 Sep -- 20 Oct & 2{,}710  & 1{,}093 & 308 & 152   \\
    \bottomrule
  \end{tabular}
  \caption{The four Telegram ecosystems: number of chats, period covered by the messages, and number of messages, forwards from other chats, \texttt{t.me} links and links to other platforms.}
  \label{tab:datasets}
\end{table}

Message histories cover approximately one month around the protests, except for chats created during that period. IIP and the original regional chats were created on 7 October. On 20 October, IIP and 14 regional chats became inaccessible. A check with a second account indicated that the loss of access was not specific to the researcher account. Administrators attributed the closure to Telegram; messages from these chats had been collected before access was lost.

To examine the split, we collected the histories of M1710 and IIP-2. M1710's history runs from its creation on 18 October to its last message on 23 October and contains 10{,}026 messages by 651 authors; the chat had 9{,}533 members. IIP-2's history covers 20--25 October and contains 7{,}433 messages by 482 authors; the chat had 2{,}325 members. We retrieved IIP-2's full member list. M1710's member list was not accessible, so identifiable users were limited to those observed posting or joining during collection.

Users were matched by Telegram identifier when comparing participation across chats. For M1710, observed users comprise accounts seen posting or joining; for IIP-2, they also include the retrieved member list. Consequently, overlap with M1710 is a lower bound. Comparisons across ecosystems use available group member lists; channel subscribers are not visible.

\subsection{Multilayer network}
\label{sec:multilayer}

We represent each ecosystem as a multilayer network~\cite{kivela2014multilayer}, with one layer per platform. Nodes in the \emph{Telegram layer} are chats. A directed edge from $u$ to $v$ indicates that $v$ forwards a message from, or links to, $u$; its weight is the number of messages establishing that relation. Nodes in the TikTok, Instagram, X, YouTube and Facebook layers are platform accounts. Inter-layer edges connect Telegram chats to the accounts they reference, weighted by the number of messages containing such links.

The network is observed from Telegram: we capture relations within its layer and links to other platforms, but no relations within those other platforms. We cannot reconstruct TikTok or Instagram follower networks, and TikTok comments lack author identifiers. A message \emph{uses} a platform layer if it contains at least one link to that platform; it uses the Telegram layer if it forwards content from another chat or contains a \texttt{t.me} link. Counting messages rather than individual links limits the influence of posts containing repeated signature links.

\subsection{Measures and annotation}
\label{sec:measures}
\label{sec:labelling}

Betweenness centrality~\cite{brandes2001faster} identifies chats that bridge the Telegram network, while the origins of forwards and links describe connections within and between ecosystems. For external platforms, \emph{coverage} is the share of chats using a platform and \emph{intensity} is the number of messages using it per 1{,}000 messages. We use coverage to compare ecosystems and intensity to compare platforms within an ecosystem. The multiplex participation coefficient~\cite{battiston2014structural} measures how broadly each chat distributes its activity across platform layers.

Messages were annotated with Claude Opus~5.5 between 19 and 25 October 2026, using Italian prompts and structured outputs. The analyses cover political orientation, stance towards the protests and promotion of violence and hate. The prompts and annotation instructions are reproduced in Appendix~\ref{app:prompts}.

\section{Findings}
\label{sec:results}

\subsection{RQ1: comparing the four ecosystems}
\label{sec:rq1}

The four ecosystems differed first in the spaces they provided for communication. Groups accounted for 29 of the 38 protest chats, including all 19 chats in the \emph{Italia in Piazza chat network}. The comparison ecosystems relied predominantly on broadcast channels: 23 of 25 conspiracy chats, 18 of 20 radical-left chats and 35 of 40 radical-right chats. Whereas these channels distributed content to subscribers, the protest groups hosted parallel conversations among participants.

The protest infrastructure was also much more recent. On 17 October, its median chat age was nine days, compared with 4.6 years for the conspiracy ecosystem, 3.6 for the radical left and 4.0 for the radical right. After the WhatsApp community reached its membership limit, a single account created IIP and 15 regional chats within 40 minutes on 7 October (Figure~\ref{fig:timeline}). Discussion was geographically distributed, but administration of the Italia in Piazza chat network was concentrated: staff accounts appeared in every accessible administrator list, and the founder administered 16 of the 19 chats, acting as the sole human administrator in eight (Figure~\ref{fig:admins}). IIP occupied the main bridging position in the protest network, with a betweenness centrality of 0.38 (Figure~\ref{fig:graphs}a; the networks of the established ecosystems are shown in Appendix~\ref{app:networks}). Other protest chats, including Udine and Naples, had independent creators and administrators.

The ecosystems also differed in where they obtained content. In the protest ecosystem, 71\% of incoming forwards and links originated within the ecosystem, compared with 82\% for the radical left, 73\% for the radical right and 26\% for conspiracy (Figure~\ref{fig:organisation}b). The conspiracy ecosystem drew on 381 external Telegram sources, compared with 98 for the protest ecosystem. Internal circulation was therefore common to several ecosystems, but in the protest network it connected recently created discussion groups rather than predominantly broadcast channels.

Connections to other platforms further distinguished these communication environments. TikTok and Instagram appeared in 50\% and 39\% of protest chats, respectively, alongside Telegram links and forwards in 63\%. TikTok sources accounted for 16\% of incoming forwards and links (Figure~\ref{fig:coverage}b--c). Conspiracy chats instead made greater use of X and YouTube, present in 60\% and 48\% of chats, while Instagram appeared in 38\% of radical-right chats. External platforms were rarely used by the radical left. Protest chats had the highest median multiplex participation coefficient, at 0.24, compared with at most 0.03 in the comparison ecosystems: their linking activity was more evenly distributed across platforms.

\begin{resultfigure}
  \centering
  \begin{minipage}[c]{0.62\textwidth}
    \centering
    \makebox[\linewidth][l]{\textsf{\small\bfseries (a)}}\par\smallskip
    \includegraphics[width=\linewidth,trim=0 0 0 14bp,clip]{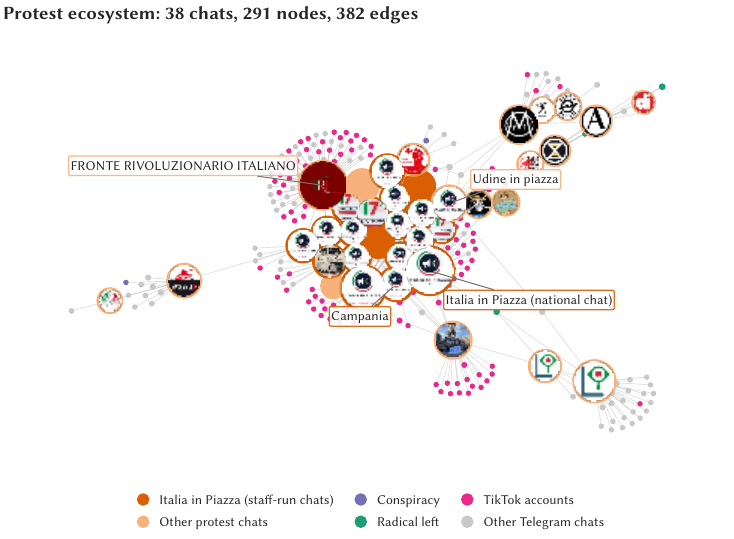}
  \end{minipage}\hfill
  \begin{minipage}[c]{0.36\textwidth}
    \centering
    \makebox[\linewidth][l]{\textsf{\small\bfseries (b)}}\par\smallskip
    \includegraphics[width=\linewidth,trim=0 0 0 14bp,clip]{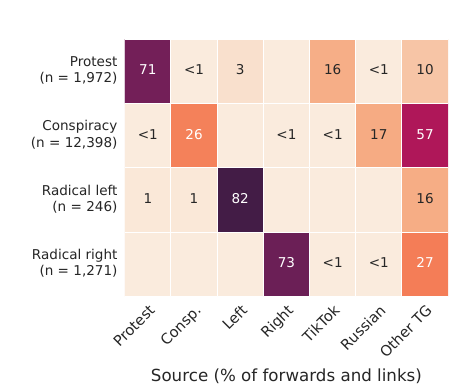}
    \par\medskip
    \makebox[\linewidth][l]{\textsf{\small\bfseries (c)}}\par\smallskip
    \includegraphics[width=\linewidth,trim=0 0 0 14bp,clip]{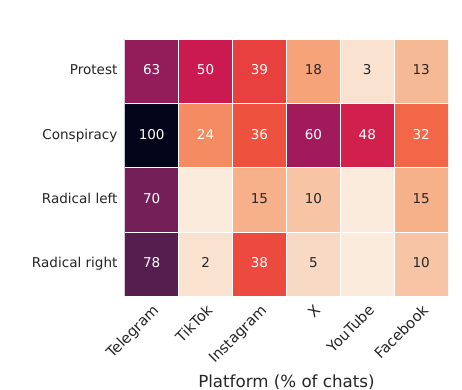}
  \end{minipage}
  \caption{Network structure and platform connections. (a) Protest chats and the sources they forward or link to; the darker ring identifies the Italia in Piazza chat network. (b) Origin of incoming forwards and links, as percentages; $n$ is the number of messages containing them. (c) Percentage of chats using each platform.}
  \label{fig:graphs}
  \label{fig:organisation}
  \label{fig:coverage}
\end{resultfigure}

These different platform profiles were accompanied by few connections between the ecosystems themselves. Forwards and links from another ecosystem accounted for at most 3\% of incoming activity. The Italia in Piazza chat network was particularly weakly connected to the comparison ecosystems: its 19 chats contained only two forwards from a radical-left channel and no observed forwards or links from conspiracy or radical-right chats. Connections involving the wider protest ecosystem came mainly from chats reached during expansion, including an Extinction Rebellion group and an anarchist circle. These connected to radical-left channels across ten chat pairs, accounting for 60 messages, and to conspiracy channels across three pairs, accounting for five messages. Across external platforms, the protest ecosystem shared no referenced TikTok or Instagram accounts with the comparison ecosystems and only three accounts on X and Facebook.

Available membership lists showed similarly limited overlap between the protest ecosystem and the established communities. Of 6{,}381 observed protest members, 11 also appeared in the conspiracy ecosystem, seven in the radical-right ecosystem and one in the radical-left ecosystem. These counts describe shared membership in the groups whose lists were accessible. Because channels predominated in the comparison ecosystems and their subscribers were not observable, the counts do not establish whether protest participants had previous affiliations with those communities.

The content of the protest chats nevertheless reflected several political positions (Figure~\ref{fig:politics}a). Among messages assigned a political orientation, 47\% were left-wing or far-left, 32\% anti-system and 6\% right-wing. The conspiracy ecosystem was more predominantly anti-system, at 69\%, with right-wing positions accounting for 15\%. The radical-left and radical-right ecosystems mainly expressed their respective orientations. The protest ecosystem thus combined positions that were more concentrated in the comparison communities, rather than expressing a single ideological profile.

\begin{resultfigure}
  \centering
  \includegraphics{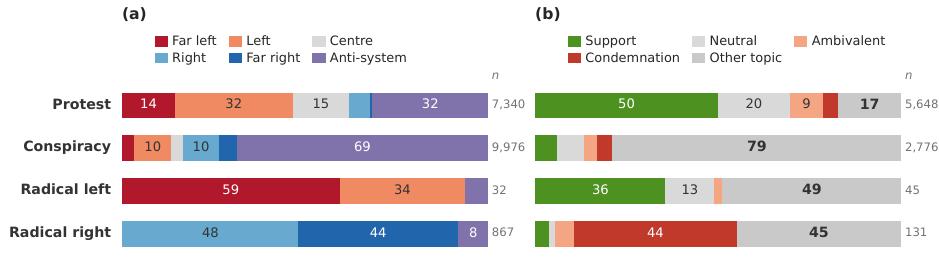}
  \caption{Positions expressed in the messages of each ecosystem, labelled by the language model. (a) Political orientation, as a share of the messages that express one. (b) Stance on the protests in the messages of 17--19 October, including those on other topics; $n$ is the number of messages.}
  \label{fig:politics}
\end{resultfigure}

Discussion of the demonstrations was concentrated in the protest ecosystem, where 83\% of messages posted on 17--19 October concerned the mobilisation (Figure~\ref{fig:politics}b). In the comparison ecosystems, the protests occupied a smaller share of discussion: 79\% of conspiracy messages, 49\% of radical-left messages and 45\% of radical-right messages concerned other topics. The demonstrations therefore remained primarily a subject of discussion within the protest chats, rather than becoming a shared focus across the established ecosystems. The radical right stood out for its predominantly negative response, with condemnation accounting for 44\% of all messages.

Unlike the comparison ecosystems, whose messages were more concentrated around particular political orientations, the protest ecosystem combined left-wing, anti-system and other positions around a shared mobilisation. Previous research on Italian Telegram similarly identified both strongly ideologically aligned communities and mixed environments brought together by a common topic, such as the war in Ukraine~\cite{alvisi2026mapping}. Here, the protests provided the common focus for a politically varied discussion space that remained largely separate from the comparison ecosystems in its connections and observed participants.

\subsection{RQ2: leadership, participants and protest practices after the split}
\label{sec:rq2}

M1710 and IIP-2 emerged from the same mobilisation but developed different leadership structures and approaches to protest. M1710 was created on the evening of 18 October. Its opening messages accused Italia in Piazza of tolerating racism and fascism and excluding participants from collective decision-making. It presented itself as antifascist and organised from below, calling for a ``permanent mobilisation'' every Saturday. IIP-2 was established two days later, after IIP and the original regional chats became inaccessible (Figure~\ref{fig:timeline}).

IIP-2 retained part of the original staff. Three of the six accounts---the vice-founder (VF) and two staff members (S1, S4)---became its administrators. Two others (S2, S3) posted in M1710 without administrative roles, while the founder's (F) account was deleted. Administrative continuity was therefore concentrated in IIP-2, whereas none of the original staff held an observed administrative role in M1710.

The reconstruction of IIP-2 also brought participants into a new central discussion space: 728 users joined within eight hours, 1{,}165 within the first day and 2{,}640 by 25 October, compared with the 4{,}073 members of IIP before its closure. Few users were observed in both successor chats: 185 accounts, approximately 7\% of each observed population. Only 44 posted in both, representing 7\% of M1710 authors and 9\% of IIP-2 authors (Figure~\ref{fig:overlap}a). These counts describe observed overlap; M1710's full member list was unavailable, so additional shared members may have gone undetected.

The chats differed in the frequency and forms of content promoting violence and hate. Under the same annotation procedure, 170 of 9{,}812 text-bearing messages in M1710 were classified in this category, compared with ten of 7{,}140 in IIP-2: 1.7\% and 0.14\%, respectively. Such messages were posted by 78 of 645 identified authors in M1710 (12\%) and eight of 474 in IIP-2 (1.7\%). The difference persisted when comparing the common observation window of 20--23 October and under the robustness checks reported in Appendix~\ref{app:validation}.

The contrast was particularly evident in messages classified as explicit calls for violence or its glorification. M1710 contained 8.7 and 5.1 such messages per 1{,}000, respectively, compared with 0.8 and zero in IIP-2. Messages proposing non-violent disruptive tactics, including blockades and occupations, were closer in frequency: 8.5 versus 5.5 per 1{,}000. These categories describe the meaning expressed in the messages, rather than determining whether the proposed actions constituted criminal offences. The difference between the chats was therefore more pronounced in language encouraging or glorifying violence than in proposals for disruptive protest (Figure~\ref{fig:incitement}b).

\begin{resultfigure}
  \centering
  \includegraphics[width=\textwidth]{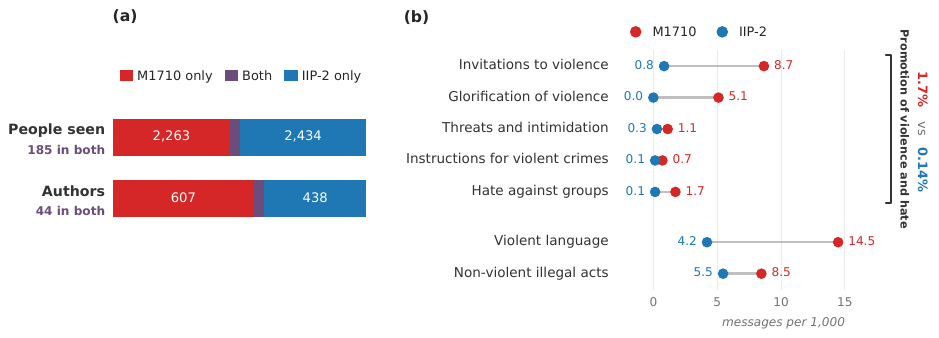}
  \caption{M1710 and IIP-2 compared. (a) Observed overlap: each row separates users seen or authors observed only in M1710, in both chats and only in IIP-2; bar lengths are proportional within each row. For M1710, only accounts observed posting or joining are identifiable, so shared membership may be underestimated. (b) Promotion of violence and hate (at least two of three annotations), messages per 1{,}000 by category. The bracket marks the five categories that form the aggregate, which covers 1.7\% of M1710 and 0.14\% of IIP-2 messages; violent language and invitations to non-violent illegal acts are recorded separately and excluded.}
  \label{fig:split}
  \label{fig:overlap}
  \label{fig:incitement}
\end{resultfigure}

Messages from IIP-2 administrators give context to this difference. When the chat opened, an administrator explicitly rejected incitement to violence as a ground rule. A statement on 21 October rejected window-smashing and isolated clashes with police as ineffective, while advocating collective, organised action. Administrators also discouraged improvised calls to gather and clarified that Italia in Piazza was not organising the demonstrations called by M1710 for 24 October. Rome and Florence administrators instead announced participation in the USB general strike of 30 October. Messages are paraphrased from Italian and attributed by role only.

The two spaces maintained different connections beyond Telegram. IIP-2 continued linking to TikTok and Instagram, at 9.6 and 4.3 messages per 1{,}000, respectively, compared with 4.9 and 3.1 in the original chats before the protests. In M1710, only 13 of 10{,}026 messages referenced either platform. M1710 encouraged migration to Signal on 21 October, deleted its city subgroups on 23 October and posted its final message that day.

The disruption following the first demonstrations exposed competing expectations about how the mobilisation should proceed. These differences took shape in two communities with markedly different norms around protest and violence: IIP-2 retained part of the original staff and explicitly promoted organised, non-violent action, whereas M1710 contained more messages encouraging or glorifying violence. The small observed overlap, including some users active in both chats, suggests that this separation did not require entirely distinct populations. Read alongside research showing that users adapt their expression to community norms~\cite{alvisi2026toxicity}, the split illustrates how a shared mobilisation can give rise to different social environments in which particular forms of expression are encouraged or discouraged. Participants may have selected spaces more compatible with their preferences, adapted to the norms they encountered, or both.

\subsection{Implications for platform security and moderation}
\label{sec:implications}

The rapid formation of the Italia in Piazza chat network shows how a single account can establish geographically distributed discussion spaces: IIP and 15 regional chats were created within 40 minutes. This combination of rapid deployment and concentrated administrative control is relevant to platform security because the same organisational capacity could be used to mobilise participants towards harmful goals. Research on disinformation as a hybrid tool highlights how information can be used to target people and critical infrastructures~\cite{alvisi2024weaponizing}. In mobilisation, the security concern is not limited to whether information is false: accurate instructions about where and when to gather could also be used to facilitate harmful action.

The closure of the original chats also shows that the loss of a discussion space need not end the activity organised around it. IIP-2 was established on Telegram within hours and attracted 1{,}165 users on its first day, while M1710 later encouraged participants to move to Signal. The cause of the original closure could not be independently established. If it resulted from a platform takedown, as administrators claimed, the subsequent reorganisation would be consistent with research showing that removals can displace communities rather than dissolve them~\cite{rogers2020deplatforming,hortaRibeiro2021platformMigrations}. The invitation to move to Signal also pointed towards communication outside the scope of our observation and with reduced visibility for platform moderation.

Reorganisation involved changes in social norms as well as discussion spaces. IIP-2 administrators explicitly rejected incitement to violence, and the chat contained ten messages classified as promoting violence or hate, compared with 170 in M1710. Read alongside research on adaptation to community norms~\cite{alvisi2026toxicity}, this contrast draws attention to the role of local administration and participant selection in shaping the environments that emerge after a split. For platform governance, the relevant outcome is therefore not only whether a chat remains accessible, but also how its participants regroup and which norms develop in the resulting spaces.

\subsection{Limitations}
\label{sec:limitations}

The study follows a single mobilisation over a short period. Loss of access to the original chats limits reconstruction of its development, and missing member lists prevent a complete account of where their participants went after closure. Comparisons rely on observable group members and activity; channel subscribers are unavailable, M1710's membership is only partially visible, and account changes cannot be resolved. Reported overlaps therefore describe the users we could identify rather than the full populations of the ecosystems.

The datasets also reflect the collection procedure. Snowball sampling favours connected chats and involves manual relevance checks, while differences between discussion groups and broadcast channels affect message volumes and comparisons across ecosystems. Connections to other platforms are observed through links shared on Telegram, not through those platforms' own interaction or follower networks.

Content labels were produced without human-coded validation. Repeated annotation checks agreement on initially flagged messages but does not measure first-pass false negatives or eliminate systematic bias. The violence-and-hate category is a definition for content analysis, not a legal assessment of the messages or their authors. Longer observation, human validation and comparison with other mobilisations would help assess how far the patterns identified here extend beyond this case.

\section{Conclusion}
\label{sec:conclusion}

The mobilisation around Italia in Piazza established a new discussion space on Telegram, with few observed connections or shared participants with the conspiracy, radical-left and radical-right ecosystems examined here. Its chats combined left-wing, anti-system and other political positions around a common focus on the protests. Within this broader ecosystem, the Italia in Piazza chat network brought together geographically distributed groups under a small administrative core, while maintaining links to TikTok and Instagram.

The disruption following the first demonstrations exposed different expectations about how the mobilisation should proceed. These took shape in two spaces with different leadership, participation and norms around protest. IIP-2 retained three of the six original staff accounts as administrators and explicitly advocated organised, non-violent action. M1710 had no original staff in observed administrative roles and contained a higher proportion of messages classified as promoting violence or hate: 1.7\%, compared with 0.14\% in IIP-2. Few users were observed in both chats, although some participated in both. The split thus involved the formation of distinct social environments within a mobilisation that had initially brought participants together.

The case shows how quickly a discussion infrastructure can be built, but also how its organisation and norms can change after a rupture. Closing the original spaces did not end the mobilisation: IIP-2 rebuilt a central chat on Telegram, while M1710 encouraged migration to Signal before ceasing to post. For platform governance, this makes the reorganisation of participants and local norms relevant alongside the availability of individual chats. The findings provide a basis for further comparisons of how online mobilisations form, split and sustain different approaches to collective action.

\section{Ethical considerations}
\label{sec:ethics}

The chats were accessible to anyone with their invitation link, but many links circulated only within the movement. Participants could therefore reasonably expect to be addressing fellow participants rather than researchers. Messages also reveal political opinions, a special category of personal data under the General Data Protection Regulation (GDPR)~\cite{gdpr2016}. We therefore do not release the dataset, including messages, membership lists, user identifiers or chat links.

Data were processed for scientific research purposes in compliance with the GDPR (Articles 5, 9(2)(j) and 89) and the Italian data protection code~\cite{dlgs196}, following the principle of data minimisation. The researcher account observed chats without posting, reacting or contacting participants. User identifiers were pseudonymised before analysis, and data are stored on institutional servers with access restricted to the authors. Text submitted to the LLM for annotation excluded author identifiers. Administrators are described by role.

\section*{Declaration on Generative AI}
During the preparation of this work, the authors used ChatGPT (OpenAI) and Claude (Anthropic) for paraphrasing, stylistic revision, and grammar and spelling checks. In addition, Claude Opus 5.5 (Anthropic) was used as a research instrument to annotate the messages analysed in this study, as described in Section~\ref{sec:labelling} and Appendix~\ref{app:prompts}. The authors reviewed and edited the AI-assisted content as needed and take full responsibility for the publication's content.

\bibliography{references}

\clearpage
\appendix

\section{Prompts}
\label{app:prompts}

The prompts were written and used in Italian, the language of the messages; we report them verbatim. Text in angle brackets is filled in for each request.

\subsection{Political orientation and stance on the protests}
\label{app:prompt-radar}

System prompt, shared by all analyses:
\begin{lstlisting}
Sei un analista di media e comunicazione politica. Analizzi i messaggi di chat Telegram (canali e gruppi) e di account TikTok per capire di cosa si parla, che posizioni emergono e da quali fonti circolano i contenuti. Valuti i contenuti, non le persone: non cercare di identificare né profilare chi scrive. Sii fattuale e bilanciato; quando un giudizio è incerto dillo. Rispondi in italiano.
\end{lstlisting}

Request for each block of messages:
\begin{lstlisting}
Argomento di interesse: Manifestazione e proteste del 17-18 ottobre 2026
Domanda dell'analisi: <domanda>

Analizza ciascuno dei <n> messaggi seguenti (il numero è tra parentesi quadre: usalo come id) e restituisci un elemento per ogni messaggio.
categoria: una tra <categorie> (scritta esattamente così); non_pertinente se il messaggio non c'entra con la domanda.
I blocchi VIDEO TIKTOK riportano il video condiviso nel messaggio: la descrizione è il contenuto che il messaggio fa circolare, i commenti sono la reazione del pubblico TikTok (non degli utenti della chat). Valuta il messaggio anche in base al video e, nella nota, segnala se i commenti mostrano una reazione diversa da quella della chat.

<messaggi, ciascuno con id, eventuale fonte del forward e testo>
\end{lstlisting}
The paragraph on TikTok videos is included only when the block contains a shared video. The two analyses use the following questions and categories:
\begin{itemize}
  \item \emph{Political orientation.} ``Quale orientamento politico esprimono i membri nei loro messaggi e commenti? Valuta la posizione che emerge dal contenuto, senza attribuirla a singole persone.'' Categories: estrema sinistra / sinistra / centro / destra / estrema destra / antisistema.
  \item \emph{Stance on the protests.} ``Che posizione hanno i contenuti verso le proteste legate all'argomento?'' Categories: sostegno / condanna / neutrale / ambivalente.
\end{itemize}
For every message the model returns, as structured output, an identifier, one to three topics, the category, a relevance flag and a short note.

\subsection{Promotion of violence and hate}

The original prompt uses the Italian word \emph{istigazione} (incitement). We report its instructions verbatim, while referring to the aggregate positive label as \emph{promotion of violence and hate} in the paper.
\label{app:prompt-incitement}

System prompt:
\begin{lstlisting}
Sei un analista che classifica messaggi di un gruppo Telegram italiano di mobilitazione politica.
Per ogni messaggio decidi se contiene ISTIGAZIONE A COMMETTERE REATI VIOLENTI O VIOLENZA (etichetta 1) oppure no (etichetta 0).

Etichetta 1 se il messaggio:
- incita o invita a commettere violenza contro persone, forze dell'ordine o cose (aggressioni, scontri, danneggiamenti, incendi, saccheggi) → categoria "incitamento_violenza"
- minaccia o intimidisce persone identificabili, anche pubblicandone o chiedendone i dati personali (nome, indirizzo, telefono, foto) per esporle a ritorsioni ("doxxing") → "minaccia_intimidazione"
- esalta o giustifica un fatto violento presentandolo come esempio da imitare o da ripetere (apologia) → "esaltazione_violenza"
- dà istruzioni operative, o condivide manuali, per commettere reati violenti o per sottrarsi all'identificazione durante azioni violente (armi, ordigni, accecare telecamere o agenti) → "istruzioni_reato"
- incita alla violenza o alla discriminazione contro un gruppo per etnia, nazionalità, religione, orientamento sessuale o genere, oppure fa propaganda d'odio contro questi gruppi anche senza un incitamento esplicito (disprezzo o insulti rivolti al gruppo in quanto tale, teorie del complotto razziste o antisemite, superiorità razziale) → "odio_discriminatorio". Una critica politica a uno Stato o a un governo (es. Israele) non è odio verso un gruppo, se non attacca il popolo o la religione in quanto tali.

Etichetta 0, categoria "nessuna": insulti e linguaggio volgare senza invito ad agire; racconti o denunce di violenze subite; discussione politica anche radicale; scioperi e manifestazioni lecite; consigli legali di autotutela (telefono spento se fermati, non parlare senza avvocato) e di protezione personale o primo soccorso (caschi, occhialini, maalox).
Etichetta 0, categoria "reato_non_violento": invito a reati NON violenti (blocchi di strade, ferrovie, porti o magazzini, occupazioni, cortei non preavvisati, resistenza passiva).
Etichetta 0, categoria "linguaggio_violento": linguaggio violento senza un invito concreto ad agire (auguri di morte o di male, iperboli come "andrebbero ammazzati", allusioni violente) da far rivedere a una persona.

Giudica il messaggio dal suo testo; "in_risposta_a" è solo contesto (il messaggio a cui risponde) e non va etichettato.
Nel dubbio tra iperbole e invito reale, guarda se c'è un invito concreto ad agire (chi, cosa, quando, dove): senza, è 0 con categoria "linguaggio_violento".
Chi riporta o cita il messaggio di altri per denunciarlo o commentarlo non istiga: giudica la posizione di chi scrive.
"motivazione": una frase breve in italiano per etichetta 1 o per le categorie "reato_non_violento" e "linguaggio_violento"; con "nessuna" stringa vuota.
Restituisci un risultato per ogni id ricevuto, nessuno in più.
\end{lstlisting}

Request for each block of 50 messages:
\begin{lstlisting}
Messaggi da classificare (uno per riga, JSON):
{"id": <id>, "testo": <testo>, "in_risposta_a": <primi 300 caratteri del messaggio a cui risponde>}
...
\end{lstlisting}
The structured output contains, for each message, the identifier, the label (0 or 1), the category and the motivation. Every message was annotated once; messages labelled 1 in this first pass were annotated twice more with the same prompts, and a message is counted as promoting violence or hate when at least two of the three annotations assign label 1.

\section{Robustness of the aggregate comparison}
\label{app:validation}

Messages promoting violence or hate are rare in IIP-2: ten messages by eight authors. Because messages from the same author are not independent, Table~\ref{tab:validation} compares several estimates with 95\% confidence intervals. The first row reports exact binomial intervals, which assume independent messages. Subsequent message-level estimates use 10{,}000 author-level bootstrap replicates, retaining all messages from each sampled author. In M1710, flagged messages are distributed across 78 identified authors, or 79 authors when posts made in the group's name are included. The most prolific author contributes 6\% of flagged messages and the ten most prolific contribute 38\%. To assess their influence, we repeat the analysis after excluding the one, three, five and ten M1710 authors with the most flagged messages. Because M1710 was active from 18 to 23 October and IIP-2 from 20 to 25 October, we also restrict both chats to their common window, 20--23 October (7{,}302 messages in M1710 and 4{,}984 in IIP-2). The final row reports the proportion of identified authors who posted at least one flagged message.

\begin{table}[h]
  \centering
  \small
  \begin{tabular}{lcc}
    \toprule
    Estimate & M1710 (\%) & IIP-2 (\%) \\
    \midrule
    Messages, exact interval            & 1.73 (1.48--2.01) & 0.14 (0.07--0.26) \\
    Messages, bootstrap over authors    & 1.73 (1.23--2.35) & 0.14 (0.05--0.26) \\
    Without the top 1 author of M1710     & 1.64 (1.18--2.26) & 0.14 (0.05--0.25) \\
    Without the top 3 authors of M1710    & 1.49 (1.09--2.00) & 0.14 (0.05--0.25) \\
    Without the top 5 authors of M1710    & 1.38 (1.01--1.86) & 0.14 (0.05--0.25) \\
    Without the top 10 authors of M1710   & 1.18 (0.86--1.60) & 0.14 (0.05--0.25) \\
    Common window, 20--23 October       & 1.47 (0.96--2.16) & 0.14 (0.03--0.29) \\
    Authors with at least one flagged message & 12.1 (9.6--14.7) & 1.7 (0.6--3.0) \\
    \bottomrule
  \end{tabular}
  \caption{Share of messages promoting violence or hate in M1710 and IIP-2, with 95\% confidence intervals. Messages: 9{,}812 in M1710 and 7{,}140 in IIP-2 (7{,}302 and 4{,}984 in the common window); identified authors: 645 and 474.}
  \label{tab:validation}
\end{table}

The intervals remain separate under every specification. M1710's lowest lower bound is 0.86\%, while IIP-2's highest upper bound is 0.29\%. The difference is therefore not driven by the days on which only one chat was active. Daily proportions show that M1710's rate was highest on 18 and 19 October (5.2\% and 2.3\%) and then remained between 1.0\% and 1.8\% from 20 to 23 October, whereas IIP-2's daily rate never exceeded 0.3\%. Author-level resampling widens M1710's interval, and excluding its most prolific authors lowers the estimated proportion, but a substantial gap remains. The author-level comparison follows the same pattern: approximately one in eight M1710 authors posted a flagged message, compared with fewer than one in fifty in IIP-2.

\section{Networks of the established ecosystems}
\label{app:networks}

\begin{figure}[htbp]
  \centering
  \includegraphics[width=0.48\textwidth]{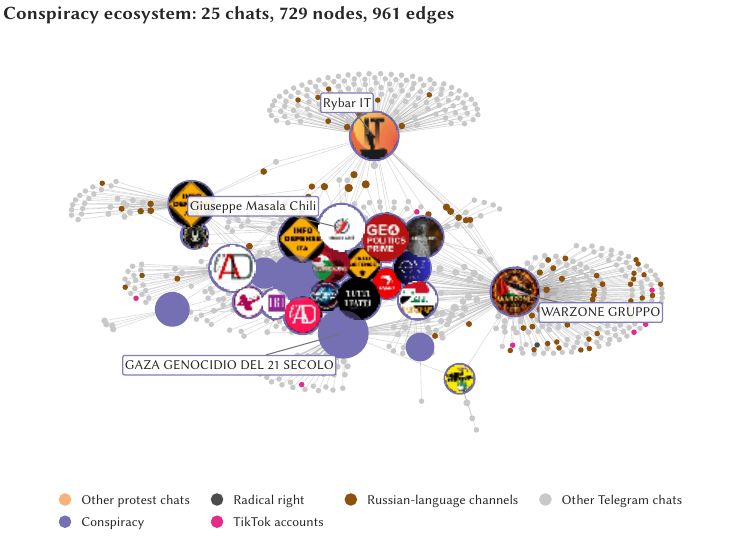}\hfill
  \includegraphics[width=0.48\textwidth]{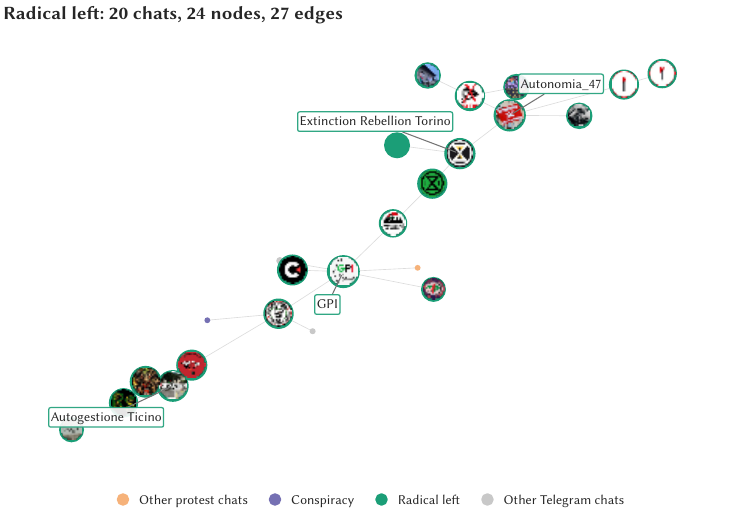}\\[8pt]
  \includegraphics[width=0.48\textwidth]{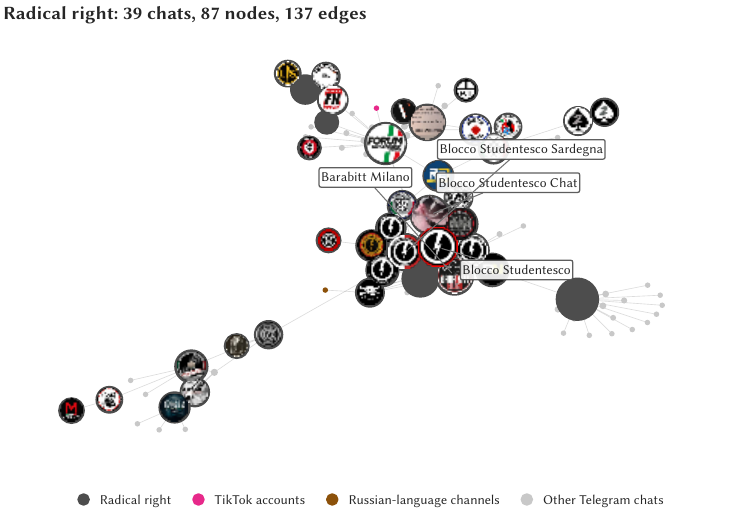}
  \caption{Telegram networks of the established ecosystems: conspiracy (top left), radical left (top right) and radical right (bottom). Nodes represent ecosystem chats and the sources they forward or link to. Recognisable profile pictures are replaced by coloured discs.}
  \label{fig:comparison-networks}
\end{figure}

Figure~\ref{fig:comparison-networks} supplements the protest network in Figure~\ref{fig:graphs}a with the corresponding Telegram networks of the three established ecosystems. The same node and source conventions apply.

\end{document}